\documentclass{article}
\usepackage{jcappub}
\usepackage[english]{babel}

\usepackage{setspace}
\usepackage{amsmath}
\usepackage{amsfonts}
\usepackage{amssymb}
\usepackage{booktabs}
\usepackage[utf8]{inputenc}
\usepackage{graphicx}
\newcommand{\nue}{\ensuremath{\nu_e}}
\newcommand{\anue}{\ensuremath{\bar{\nu}_e}}
\newcommand{\numu}{\ensuremath{\nu_\mu}}
\newcommand{\anumu}{\ensuremath{\bar{\nu}_\mu}}
\newcommand{\nutau}{\ensuremath{\nu_\tau}}
\newcommand{\anutau}{\ensuremath{\bar{\nu}_\tau}}
\newcommand{\tht}[1]{\ensuremath{\theta_{#1}}}
\newcommand{\dcp}{\ensuremath{\delta_{CP}}}
\newcommand{\edep}{\ensuremath{E_\mathrm{dep}}}
\newcommand{\enu}{\ensuremath{E_\nu}}
\newcommand{\mean}[1]{\ensuremath{\langle #1 \rangle}}

\renewcommand{\deg}{\ensuremath{^\circ}}
\usepackage{newunicodechar}
\newunicodechar{σ}{\ensuremath{\sigma}}
\newcommand{\Gen}{IceCube-Gen$2$}

\newcommand{\onehalf}{\ensuremath{\frac{1}{2}}}
\newcommand{\onethird}{\ensuremath{\frac{1}{3}}}
\newcommand{\twothird}{\ensuremath{\frac{2}{3}}}

\title{Energy-dependent flavour ratios in neutrino telescopes from charm}

\author[a]{Atri Bhattacharya,}
\affiliation[a]{Space sciences, Technologies and Astrophysics Research (STAR)
Institute, Université de Liège, Bât.~B5a, 4000 Liège, Belgium}

\author[b]{Rikard Enberg,}
\affiliation[b]{Department of Physics and Astronomy, Uppsala University, Box
516, 751 20 Uppsala, Sweden}

\author[c]{Mary Hall Reno}
\affiliation[c]{Department of Physics and Astronomy, University of Iowa, Iowa
City, IA 52242, USA}

\author[d,e]{and \\Ina Sarcevic}
\affiliation[d]{Department of Physics, University of Arizona, Tucson, AZ 85721,
USA}
\affiliation[e]{Department of Astronomy and Steward Observatory, University of
Arizona, Tucson, AZ 85721, USA}

\emailAdd{A.Bhattacharya@uliege.be}
\emailAdd{rikard.enberg@physics.uu.se}
\emailAdd{mary-hall-reno@uiowa.edu}
\emailAdd{ina@physics.arizona.edu}

\abstract{The origin of the observed diffuse neutrino flux is not yet known.
	Studies of the relative flavour content of the neutrino flux detected at Earth
	can give information on the production mechanisms at the sources and on flavour
	mixing,  complementary to measurements of the spectral index and normalization.
	Here we demonstrate the effects of neutrino fluxes with different spectral
	shapes and different initial flavour compositions dominating at different
	energies, and we study the sensitivity of future measurements with the IceCube
	Neutrino Observatory. Where one kind of flux gives way to another, this shows
	up as a non-trivial energy dependence in the flavour compositions. We explore
	this in the context of slow-jet supernovae and magnetar-driven supernovae---two
	examples of astrophysical sources where charm production may be effective.
	Using current best-fit neutrino mixing parameters and the projected 2040
	IceCube uncertainties, we use event ratios of different event morphologies at
	IceCube to illustrate the possibilities of distinguishing the energy dependence
	of neutrino flavour ratios.}

\begin{document}
\maketitle

\section{\label{sec:intro}Introduction}
%!TEXroot = ./main.tex

Our understanding of high energy neutrinos at PeV energies and higher and their sources
has been bolstered by the stream of data collected at the
IceCube Neutrino Observatory (IC)~\cite{IceCube:2002eys} since
2011, first published in Refs.~\cite{IceCube:2013cdw,IceCube:2013low}.
Analyses of IceCube data have repeatedly confirmed the existence of neutrino
fluxes higher than---and extending to energies beyond---atmospheric
neutrinos~\cite{IceCube:2018pgc,Stettner:2019tok,IceCube:2020wum,IceCube:2020acn}.
Conventionally, one expects such highly energetic neutrinos to come from
extra-galactic astrophysical sources such as Active Galactic Nuclei (AGNs) or
Gamma-Ray Bursts (GRBs), where they are produced due to collisions of rapidly
accelerated protons with protons or with the copious abundance of  photons in
such environments.
Alternatively, beyond the Standard Model (BSM) mechanisms such as the decay or
annihilation of very heavy PeV-scale dark matter, may be responsible for the
production of ultra-high energy neutrinos~(see,
e.g.,~\cite{Bhattacharya:2014vwa,Bhattacharya:2014yha,Chianese:2017nwe,
	Chianese:2019kyl,Skrzypek:2022hpy}).\footnote{To clarify our notation, we use
	the term \emph{ultra-high energy} neutrinos to refer to neutrinos with energies
	$E \geq 100$ TeV.}
Further, the total flux may  be a combination of astrophysical and BSM
neutrino fluxes with different relative normalisation.
More data in the future, potentially aided by a bigger version of IceCube
called \Gen\ \cite{IceCube-Gen2:2020qha}, will be needed to effectively
determine their origins precisely.

The extra-terrestrial neutrino flux at IceCube is inferred to be nominally
consistent with a uniform power-law~\cite{IceCube:2018pgc,Stettner:2019tok,IceCube:2020wum,IceCube:2020acn}:
\begin{equation}
	E^2 \frac{d\Phi}{dE} = \Phi_0 E^{-\gamma + 2}.
	\label{eq:icflux}
\end{equation}
There are hints, albeit inconclusive, due to the low event statistics
typical of high energy neutrino fluxes, of disagreements
between inferences of the flux spectrum from different data samples.
For example, analysis of high energy starting events (HESE) collected
over 7.5 years points to a steeply falling (or soft) spectrum,
with the best-fit value of $\gamma = 2.88$~\cite{IceCube:2020wum}.
An analysis involving only through-going muons seen over 9.5 years points to
a much harder spectrum, with $\gamma = 2.3$~\cite{Stettner:2019tok}.
Other analyses with different subsets of IceCube data find best-fit values for
the spectral index lying between these
extremes~\cite{IceCube:2018pgc,IceCube:2020acn}.
Several solutions discussing how these discrepancies may be ameliorated have
been suggested; one among them is to assume that, at the higher end of the
spectrum, neutrino fluxes from choked jets may
dominate~\cite{Bhattacharya:2014sta,IceCube:2020acn}.

No matter their origins, a fraction of these neutrinos arrive at Earth after having
traversed large cosmological distances.
Along the way, the three flavours of neutrinos, \nue, \numu, and \nutau,
oscillate amongst themselves leading to a redistribution of their relative
flavour content by the time they are detected at Earth vis-à-vis at the
source~\cite{Learned:1994wg,Pakvasa:2007dc}.
Therefore, complementary to studies of the shape and normalisation of the total
flux, analyses of the relative neutrino flavour content as they appear in the
neutrino flux at Earth can throw light on the mechanism of their production at
source and on the physics of flavour mixing during
propagation~\cite{Kachelriess:2006ksy,Choubey:2009jq,Hummer:2010ai,Mena:2014sja,Bustamante:2019sdb,Song:2020nfh,Riabtsev:2022ynm,Arguelles:2023bxx,Testagrossa:2023ukh}.
The former helps to understand the source environment where these interactions
take place while the latter may be able to pinpoint traces of BSM physics that
influence neutrino oscillation (see~\cite{Ahlers:2018mkf} for a review and
references therein).

The IceCube Neutrino Observatory, with its three-dimensional arrangement of photo-detectors distributed
over a large volume, is capable of detecting different event morphologies
arising from different neutrino flavours interacting with nucleons in rock/ice:
(a) cascades with their starting vertex inside the instrumented volume,
initiated either by  charged-current deep-inelastic scattering (DIS) involving
electron or low-energy tau neutrinos with nucleons ($N$), or by neutral-current
DIS of any neutrino flavour;
(b) tracks from charged-current interaction of a \numu\ and a nucleon;
or,
(c) at energies more than a PeV, tau neutrino specific signatures such as
double bangs from $\nutau N$ charged-current interactions shortly followed by
$\tau$ decay.\footnote{At low energies below $\lesssim$ 1 PeV these are
difficult to distinguish from $\nue$ charged-current cascades owing to the
$\tau$ decay happening very close to the initial $\nutau N$ interaction.}
This makes it the only operating neutrino detector that can measure individual
flavour components in the incoming extra-galactic neutrino flux, within the
limits defined by the above event morphologies. In the future,
Baikal-GVD~\cite{Eckerova:2023ynw} and KM3NeT/ARCA~\cite{vanEeden:2023ucx},
both currently being built, may also be able to measure the  flavour
composition of astrophysical fluxes.

Inferring neutrino flavour composition at the source from that obtained from
cascade, track, and tau relative event rates at IceCube requires knowledge of
the neutrino mixing parameters from precision neutrino oscillation experiments
at Earth.  Indeed the ever narrowing error bars on the measurements of mixing
angles \tht{12}, \tht{13}, and \tht{23} at experiments like
Super-Kamiokande~\cite{Super-Kamiokande:2017yvm}, Sudbury Neutrino
Observatory~\cite{SNO:2002tuh}, Daya Bay~\cite{DayaBay:2012fng}, etc.,\ have
gone a long way towards improving the robustness of such inferences.

Using these inputs, several studies in the past have studied how to infer the
flavour composition using relative event rates of cascade and track
morphologies~\cite{Mena:2014sja,IceCube:2015rro,Bustamante:2019sdb}.
These typically assume that the flux composition remains unchanged across the
entire range of ultra-high energies $10$ TeV--$10$ PeV where IceCube sees data.

However, it may be the case that neutrino fluxes originating from fundamentally
different interactions---and thus with different spectral shapes and different
initial flavour compositions---dominate at different energy ranges.
If two or more competing fluxes indeed have different overall shapes and
initial flavour compositions between 10 TeV and 10 PeV, the
transition from one flux to another will induce an energy dependence in the
flavour compositions seen at IceCube.
An analysis of flavour composition assuming an unchanged flux over the
entire range of energy will have missed this transition.

In this work we explore the impact of such a possibility.
As an example, we consider the case of diffuse neutrino fluxes from slow-jet
supernova (SJS) sources, where under the right conditions, neutrino fluxes from
pion and kaon decays, i.e.,\ with an initial relative composition of $\nue +
\anue : \numu + \anumu : \nutau + \anutau = \onethird :\twothird :0$, dominate
up to energies of about 100 TeV, giving way to neutrinos from decays of heavier
charmed mesons, the latter produced with the flavour composition $\onehalf
:\onehalf :0$.
We consider the case of magnetar-driven supernovae where a similar
transition takes place at much higher energies $\sim 10^{10}$ GeV, with
potential consequences for flavour composition studies at \Gen\ in the future.
We also evaluate an example in which the ultrahigh-energy neutrino flux comes
from a mix of $\pi/K$ and charm decays to neutrinos in sources.

We note that energy-dependent flavour composition transitions may also
be caused by exotic Beyond Standard Model (BSM) physics effects such as neutrino
decay or Lorentz invariance violation acting during neutrino propagation (see
Refs.~\cite{Ohlsson:2012kf} and~\cite{Ahlers:2018mkf} for recent reviews
and references therein).
Some of these BSM effects may end up enforcing flavour ratio
transitions similar, at least qualitatively, to those coming from the
dominance of charm decay initiated neutrino fluxes at higher energies
(as discussed in this work).
This would make it difficult to disentangle the different underlying mechanisms.
However, a full analysis involving such BSM effects is beyond the scope of this
work.

The next section describes the two astrophysical sources of fluxes of neutrinos
from charm. Section~\ref{sec:nuosc} reviews the effect of neutrino oscillations
on flavour compositions at Earth, given source flavour ratios, as a function of
energy. Our analysis of event-rate ratios based on event morphologies from
different neutrino flavour induced events in IceCube is described in
Section~\ref{sec:ic}. Our summary and conclusions appear in
Section~\ref{sec:conclu}.

\section{\label{sec:fluxes}Neutrino fluxes at source from charm}
%!TEXroot = ./main.tex

In this section, we provide a brief description of the different sources of
neutrino fluxes we consider for our case studies. We refer to
Refs.~\cite{Enberg:2008jm,Bhattacharya:2014sta,Carpio:2020wzg} for detailed
descriptions of the models that we use.

We consider two main sources of neutrinos:\ slow-jet supernovae (SJS) and
magnetar-driven supernovae (MdSn), which have in common that they yield a
contribution of neutrinos produced from decays of charm quarks, which has both
a different flavour composition and a different energy dependence than the
conventional neutrinos produced from decays of pions and kaons. We stress that
these are two examples of such sources that we use as simple models to point
out the possible mechanism, but there may be other types of sources with other
energy dependencies that give this effect too.

\begin{figure}
	\centering
	\includegraphics[width=0.75\textwidth]{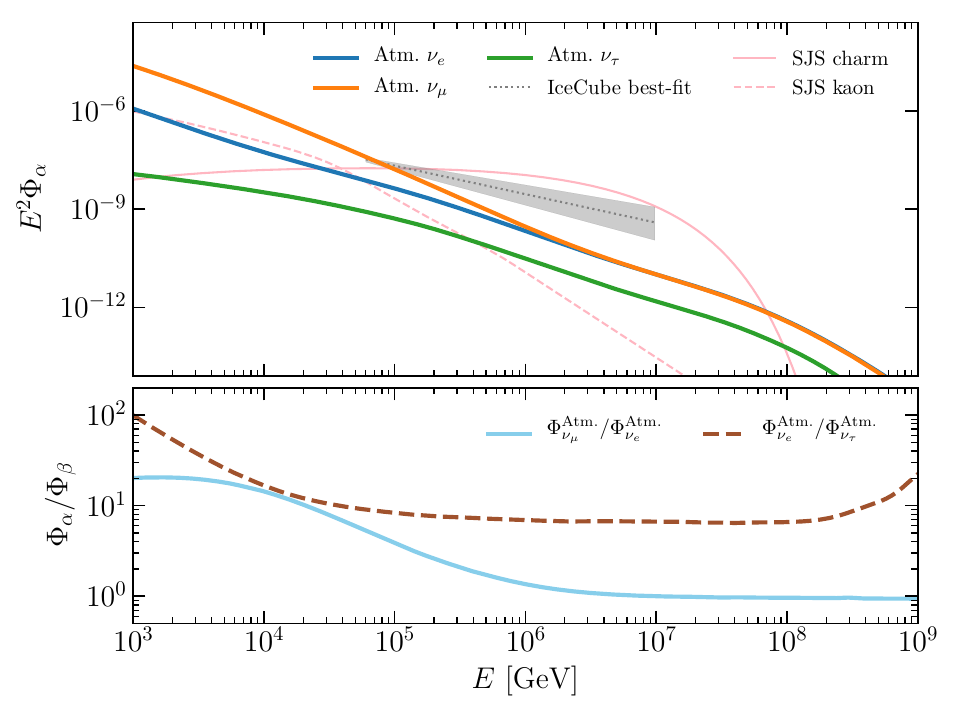}
	\caption{\label{fig:fluxes} Angle averaged atmospheric neutrino
		fluxes~\cite{Fedynitch:2015zma} used to compute flavour ratio variations in
		Figs.~\ref{fig:flvar}, \ref{fig:flvarscaled} and \ref{fig:SJSC2T}. Also shown are SJS charm and kaon diffuse
		$\nu_\mu+\bar\nu_\mu$ fluxes from Ref.~\cite{Bhattacharya:2014sta} as used
		in this work, as is the best-fit uniform power-law flux (per flavour)
		inferred from 7.5-yr high-energy starting events data at
		IceCube~\cite{IceCube:2020wum}.
		The bottom panel shows the non-trivial flavour ratios, unrelated to
		neutrino oscillation, for the atmospheric fluxes as a function of energy.}
\end{figure}

\subsection{Slow-jet supernova sources}

We first consider slow-jet supernovae
(SJS)~\cite{Razzaque:2004yv,Razzaque:2005bh,Ando:2005xi}, where the neutrino
flux is generated in a relativistic jet launched by a core-collapse supernova.
The neutrino generation mechanism is similar to the standard scenario for
gamma-ray bursts~\cite{Waxman:1997ti}: Protons are accelerated by Fermi
acceleration in the jet and collide with ambient photons, and neutrinos are
generated from decays of pions and kaons produced in the $p\gamma$ collisions,
provided that the centre-of-mass energy exceeds the threshold for $\Delta^+$
production. Moreover, in a GRB the jet is ultra-relativistic and emits a burst
of gammas generated by synchrotron radiation and inverse Compton scattering.

In contrast, the jet in an SJS is mildly relativistic. It is choked and stalls
before it punches through the envelope, so there is no accompanying gamma ray
burst, but there will be a visible optical supernova. But because of the
different environment with a larger optical depth, in an SJS there may
additionally be $pp$ collisions that produce pairs of charm quarks, which, when
they decay,  give rise to an additional component of the neutrino flux.
The relative sizes of the $pp$ and $p\gamma$ components depend on the
characteristics of the astrophysical environment (see
also~\cite{Murase:2013ffa}).

The $pp$ production of charm quarks has another important consequence. Pions
and kaons are long-lived and will undergo both hadronic and radiative energy
losses, or cooling, before decaying, so that the neutrino energies are
downgraded. However, for the charmed hadrons ($D$ mesons and $\Lambda_c$
baryons) that are produced when the charm quarks hadronise, there are two
factors to contribute to making them have much less cooling:\ First, they are
very short-lived and decay before they interact to lose energy from hadronic
cooling. Second, the efficiency of radiative cooling  scales as (mass)$^{-4}$,
meaning that the heavier charmed hadrons are cooled less than pions and kaons.

On the other hand the production cross section for charmed hadrons is several
orders of magnitude smaller than the cross section for pions and kaons, but in
the competition between production and cooling it turns out that neutrinos from
charm decays may dominate the neutrino flux at high energies. The cross-over
energy where this happens depends on the detailed properties of the
astrophysical source and the maximum energy of accelerated protons in the jet.
For example, it turns out that in AGNs, the cross-over energy is at a higher
energy than the maximum proton energy making this effect non-existent, but in
SJS it may be an important effect~\cite{Enberg:2008jm}.

In  Refs.~\cite{Enberg:2008jm,Bhattacharya:2014sta} we demonstrated that for
realistic SJS properties there may be an appreciable component of neutrinos
from charm decay that may be an important contribution to the diffuse flux
detected by the IceCube neutrino observatory.
In~\cite{Valtonen-Mattila:2022nej} it was further argued that \Gen\ has
capabilities to detect SJS sources out to megaparsec distances.

\subsection{Magnetar-driven supernovae}

Magnetars are strongly magnetised, fast-rotating neutron stars that are
expected to be sources of high energy neutrinos.  Cosmic rays are likely
accelerated to ultra-high energies by magnetic dissipation of the spindown
energy in the magnetar, and do not escape the supernova ejecta, but rather
produce charged pions through $pp$ and/or $p\gamma$ interactions, which then
decay to high-energy neutrinos.  In Ref.~\cite{Carpio:2020wzg}, only the $pp$
interactions are included since they dominate. In addition to pion production
and decay, there are kaons and charmed hadrons that are produced that decay
into neutrinos. As for SJS, neutrino production from charm is often neglected
but can become the dominant contribution to the neutrino flux at early times
when pions and kaons experience significant cooling before they decay. In
magnetars, this significant cooling due to interactions of pions and kaons with
the surrounding protons and photons happens for energies above $10^9$ GeV.
Thus for these energies, prompt decays of charm hadrons can be dominant source
of very high energy neutrino production.

In Ref.~\cite{Carpio:2020wzg}, the neutrino fluence was evaluated for
magnetar-driven supernovae (MdSn) and for magnetar-driven merger novae.  In the
former case, the supernova ejecta mass was taken to be between $10$ and $35$
solar masses ($M_\odot$), and the nucleon density was assumed to be
homogeneous. For the latter case, when the rapidly rotating magnetars are born
at the merger of the low-mass neutron star binaries, ejecta masses are taken to
be between $10^{-2} M_\odot$ and $10^{-1} M_\odot$.
While the charm contribution is not dominant over the full energy range,
charm becomes significant for energies above $\sim 10^9$ GeV at the beginning
of the burst. Late time emission is dominated by pion and kaon contributions.

Magnetar-driven supernovae with a magnetic field $B=10^{14.5}$ G, an initial
angular frequency of $10^{-4}$ s and the rate of $500$ yr$^{-1}$ Gpc$^{-3}$
have pion dominated neutrino fluxes for energies up to $\sim 4 \times 10^9$
GeV, at which point kaon decays into neutrinos become important until $E_\nu
	\sim  10^{10}$ GeV  when charm takes over~\cite{Carpio:2020wzg}. The flavour
ratio can be extracted from these fluxes as a function of energy.
The overall uncertainty in the prediction of the neutrino flux from charmed
meson decay is about an order of magnitude.  Here, we use the central values of
the prediction of the neutrino flux from charm.

\subsection{Atmospheric neutrinos}

An additional component of the detected neutrino flux, which is a background to
the sources discussed above, is the atmospheric neutrino flux. Interactions of
cosmic rays with the Earth atmosphere produce pions, kaons and charmed mesons.
These mesons decay and produce neutrinos; pions and kaons give muon and
electron neutrinos and antineutrinos, and secondary neutrinos are produced from
muon decays, where muons come from pions and kaons. On the other hand, charmed
mesons decay into all three flavours of neutrinos, thus producing $\nu_\tau$
neutrinos, albeit to a lesser extent than $\nu_e$ and
$\nu_\mu$~\cite{Pasquali:1998xf,Bhattacharya:2016jce}.
A direct flux of $\nu_\tau+\bar\nu_\tau$ cannot come from pion and kaon decays.
The atmospheric neutrino flux has been measured up to energies of several PeV
by IceCube, currently without evidence for prompt neutrinos.

We compute the total atmospheric neutrino flux for each of the three flavours
using \verb|MCEq|~\cite{Fedynitch:2015zma}, including both the conventional
component from pions and kaons and the prompt component from charm.  The former
is computed based on Hillas-Gaisser's 2012 \texttt{H3a} cosmic-ray
model~\cite{Gaisser:2011klf} and the \texttt{SIBYLL-2.3c} hadronic interaction
model~\cite{Riehn:2017mfm}, while the latter is based on
Ref.~\cite{Martin:2003us}.  In Fig.~\ref{fig:fluxes} we show the angle averaged
atmospheric flux for different flavours, as well as the diffuse neutrino flux
from slow jet supernovae.  The IceCube best fit for high energy neutrinos from
ref.~\cite{IceCube:2020wum} sits only slightly below the predicted SJS
neutrinos from charm.  The lower plot shows that at higher energies, above a
PeV, the atmospheric \nue\ flux becomes identical to the \numu\ flux, both
about an order of magnitude bigger than the \nutau\ component.  The equality of
the \nue\ and \numu\ flux for $E\gtrsim 10^7$ GeV is a consequence of the charm
contribution becoming more important at higher energies.

High energy atmospheric $\nu$ fluxes are not affected by neutrino oscillation,
as these travel relatively short distances (up to the diameter of the Earth)
before they are detected; their relative compositions are determined solely by
the underlying production mechanism. This is in contrast to low energy
atmospheric neutrinos. In their traverse of as much as the Earth's diameter
through rock, the three flavours oscillate~\cite{Wallraff:2014qka}.  Indeed,
such neutrinos are used by experiments such as Super-Kamiokande to determine
oscillation parameters~\cite{Super-Kamiokande:2017yvm}.

\subsection{Other high energy neutrino fluxes}

The cut-off in the cosmic ray spectrum at cosmic ray energies of $\sim 50$ EeV
could be a signal of Greisen-Zatsepin-Kuzmin (GZK)
effect~\cite{Greisen:1966jv,Zatsepin:1966jv} in which the cosmic ray-photon
interaction is above the threshold for $\Delta^+$ production. Whether or not
this is the GZK effect or a signal of the limits of cosmic accelerators, cosmic
ray interactions with the cosmic background photons will produce a diffuse flux
of ultra-high energy
neutrinos~\cite{Berezinsky:1969erk,Stecker:1978ah,Hill:1983mk,Yoshida:1993pt,
	Engel:2001hd,Kotera:2010yn,Ahlers:2012rz,Heinze:2015hhp,AlvesBatista:2018zui,
	Heinze:2019jou}.
With the GZK effect, this cosmogenic neutrino flux has primarily $\bar\nu_e$
that come from neutron decays for $E_\nu\lesssim 100$ PeV, while at higher
energies, the initial flavour ratios are $\onethird : \twothird :0$
characteristic of pion production and decay, along with secondary muon decays.
While there is an interesting evolution of flavour ratios as a function of
energy in the cosmogenic neutrino flux, charm production is unlikely to be a
feature in the flavour composition of cosmogenic neutrinos since the cosmic
ray-photon cross section for charm-pair production is significantly smaller
than for $\Delta$ production~\cite{Goncalves:2006ch}.

\paragraph{}
Beyond Standard Model physics at the source could, in principle, involve charm
production and decays, for example in the case of super-heavy dark matter
annihilation to charm pairs (see,
e.g.,~\cite{Esmaili:2013gha,Chianese:2016kpu,Bhattacharya:2017jaw,
Bhattacharya:2019ucd}).
This annihilation channel could give interesting neutrino flavour signatures. A
complication in the neutrino flavour ratios as a function of energy is that the
charm states produced are highly virtual, so they will shower, hadronise and
decay~\cite{Bauer:2020jay}. Different from charm production in sources, the
dark matter annihilation essentially occurs in vacuum, so cooling processes do
not come into play and all of the hadrons decay. The energy dependent neutrino
flavour composition of the resulting neutrino flux is beyond the scope of the
present work.

\section{\label{sec:nuosc}The effect of neutrino oscillation on flavour
  compositions at Earth}%
%!TEXroot = ./main.tex

When neutrinos originate at distant, extra-galactic astrophysical sources such
as GRBs, AGNs, SJS, etc., owing to the large distances the neutrinos
subsequently propagate before reaching Earth, the oscillatory terms in the
usual neutrino oscillation probability average out.
As a consequence, the three flavours of neutrinos mix with each
other incoherently with the probability~\cite{Learned:1994wg,Pakvasa:2007dc}:
\begin{equation}
	P_{\nu_\alpha \to \nu_\beta} = \sum_{k=1,2,3}
	\lvert U_{\alpha k} \rvert^2
	\lvert U_{\beta k} \rvert^2
	\label{eq:nuosc}
\end{equation}
describing the transition of flavour $\nu_\alpha$ to $\nu_\beta$.
Here, $U_{ij}$ represents the $ij$-th element of the so-called PMNS matrix that describes
the mixing between flavour and mass eigenstates of neutrinos~\cite{ParticleDataGroup:2022pth}:
\begin{align}
	U_\text{PMNS}
	 & =
	\begin{bmatrix} 1 & 0       & 0      \\
                0 & c_{23}  & s_{23} \\
                0 & -s_{23} & c_{23}
	\end{bmatrix}
	\begin{bmatrix} c_{13}                       & 0 & s_{13}e^{-i\delta_\text{cp}} \\
                0                            & 1 & 0                            \\
                -s_{13}e^{i\delta_\text{cp}} & 0 & c_{13}
	\end{bmatrix}
	\begin{bmatrix} c_{12}  & s_{12} & 0 \\
                -s_{12} & c_{12} & 0 \\
                0       & 0      & 1
	\end{bmatrix} \\
	 & =
	\begin{bmatrix} c_{12}c_{13}                                            & s_{12} c_{13}   & s_{13}e^{-i\delta_\text{cp}} \\
                -s_{12}c_{23} - c_{12}s_{23}s_{13}e^{i\delta_\text{cp}} & c_{12}c_{23}
                - s_{12}s_{23}s_{13}e^{i\delta_\text{cp}}               & s_{23}c_{13}                                   \\
                s_{12}s_{23} - c_{12}c_{23}s_{13}e^{i\delta_\text{cp}}  & -c_{12}s_{23} -
                s_{12}c_{23}s_{13}e^{i\delta_\text{cp}}                 & c_{23}c_{13}
	\end{bmatrix}.
	\label{eq:upmns}
\end{align}
where $ c_{ij}$ and $ s_{ij}$ are used to denote the
$\cos(\tht{ij})$ and $\sin(\tht{ij})$ of the mixing angles $\theta_{ij}$
respectively.
The CP-violating phase in represented as \dcp.

The increasingly precise measurements of the neutrino mixing matrix are thanks
to solar, atmospheric, and reactor neutrino data collected by dedicated
neutrino detectors~\cite{SNO:2002tuh,KamLAND:2008dgz,Super-Kamiokande:2017yvm,
	DayaBay:2012fng,DoubleChooz:2011ymz,RENO:2012mkc,K2K:2006yov,
	MINOS:2020llm,T2K:2023smv,NOvA:2019cyt,T2K:2018rhz}.
In particular, recent studies of reactor neutrino data at Daya
Bay~\cite{DayaBay:2012fng}, Double Chooz~\cite{DoubleChooz:2011ymz}, and
RENO~\cite{RENO:2012mkc} have significantly narrowed the uncertainties on the
\tht{13}.  Likewise, precision information on \tht{12} mainly comes from solar
neutrino experiments, while atmospheric neutrino experiments and muon neutrino
disappearance searches at Long Baseline experiments such as
K2K~\cite{K2K:2006yov}, MINOS~\cite{MINOS:2020llm}, T2K~\cite{T2K:2023smv}, and
NO$\nu$A~\cite{NOvA:2019cyt} throw light on \tht{23} specifically.  Finally,
the CP violating phase \dcp\ is only weakly constrained, primarily from T2K
data~\cite{T2K:2018rhz}.

Improved precision on these parameters, has in turn allowed us to understand
interesting facets of neutrino phenomenology,
including deviations from tri-bi-maximal mixing, the \tht{23} octant, and hints
towards the true mass ordering.

\subsection{Best-fit parameters and uncertainties}

For our analysis, we use the most recent mixing parameters as tabulated in
Nu-Fit~\cite{Esteban:2020cvm,nufit-web} obtained when using atmospheric SK data
and assuming normal mass hierarchy.
We find no tangible change in results when using best-fit parameters obtained
assuming inverted hierarchy instead.

The 2040 projections on parameter uncertainties are obtained from
Refs.~\cite{DayaBay:2012fng,JUNO:2022mxj,Hyper-Kamiokande:2018ofw,DUNE:2020lwj},
and listed in Table~\ref{tab:oscparams} (see also Ref.~\cite{Denton:2022een}).
In the analysis below, we assume that the best-fit values of these parameters
remain the same as in current data.

\begin{table}[htb]
	\centering
	\begin{tabular}{l@{\hspace{2cm}} r@{\hspace{2cm}} r}
		\toprule
		Parameter        & Current Best-fit             & 2040 uncertainty ($\pm$)
		\\
		\midrule
		$\sin^2\tht{13}$ & $0.0223_{-0.0018}^{+0.0017}$ &
		0.0018~\cite{DayaBay:2012fng}
		\vspace{0.5em}
		\\
		$\sin^2\tht{12}$ & $0.303_{-0.0330}^{+0.0338}$  &
		0.005~\cite{JUNO:2022mxj}
		\vspace{0.5em}
		\\
		$\sin^2\tht{23}$ & $0.451_{-0.042}^{+0.153}$    &
		0.018~\cite{Hyper-Kamiokande:2018ofw}
		\vspace{0.5em}
		\\
		\dcp\            & $232{\deg_{~}}_{-88}^{+118}$ &
		35\deg\ \cite{DUNE:2020lwj}
		\\
		\bottomrule
	\end{tabular}
	\caption{\label{tab:oscparams}Oscillation parameters relevant to UHE flavour
		mixing and their current and 2040 projected 3$\sigma$ uncertainties. Note
		that the 2040 projected $3\sigma$ uncertainty on $\theta_{23}$ has been
		scaled up from the $1\sigma$ data specified in
		Ref.~\cite{Hyper-Kamiokande:2018ofw}, and should be considered
		approximate.}
\end{table}

\subsection{Flavour composition at Earth}

Using the values of the mixing matrix parameters as determined at neutrino
oscillation experiments at Earth, it becomes straightforward to use
equation~\ref{eq:nuosc} to compute the flavour composition of the neutrino flux
at Earth, $\Phi_\alpha$, given an initial composition at source.

To quantify the flavour composition of the neutrino flux at Earth, we look at
different measures:
\begin{itemize}
	\item
	      flavour fractions defined as
	      \begin{equation}
		      f_\alpha = \frac{\Phi_\alpha}{\Phi_\mathrm{total}}
		      = \frac{\Phi_\alpha}{\sum\limits_{\beta}\Phi_\beta}\,,
		      \label{eq:flcomp}
	      \end{equation}
	      where the subscripts represent neutrino flavours and the sum over
	      $\beta$ runs over all three to give the total flux;
	\item
	      flavour ratios between the fluxes of any two flavours $\nu_\alpha$ and
	      $\nu_\beta$ of the three comprising the total flux, defined as
	      \begin{equation}
		      r_{\alpha\beta} = \frac{\Phi_\alpha}{\Phi_\beta}
		      = \frac{f_\alpha}{f_\beta}\, ,
		      \label{eq:flrat}
	      \end{equation}
	\item
	      and finally, ratios of cascades to $\mu$-tracks and high-energy \nutau\
	      double bangs, considered in a later section.
\end{itemize}
For each measure, the best-fit values of PMNS mixing matrix parameters give us the
central values, while the corresponding upper and lower limits are
obtained by varying oscillation parameters within their allowed 3$\sigma$
ranges, and determining the relevant maxima and minima respectively.

\begin{figure}
	\centering
	\includegraphics[width=0.75\textwidth]{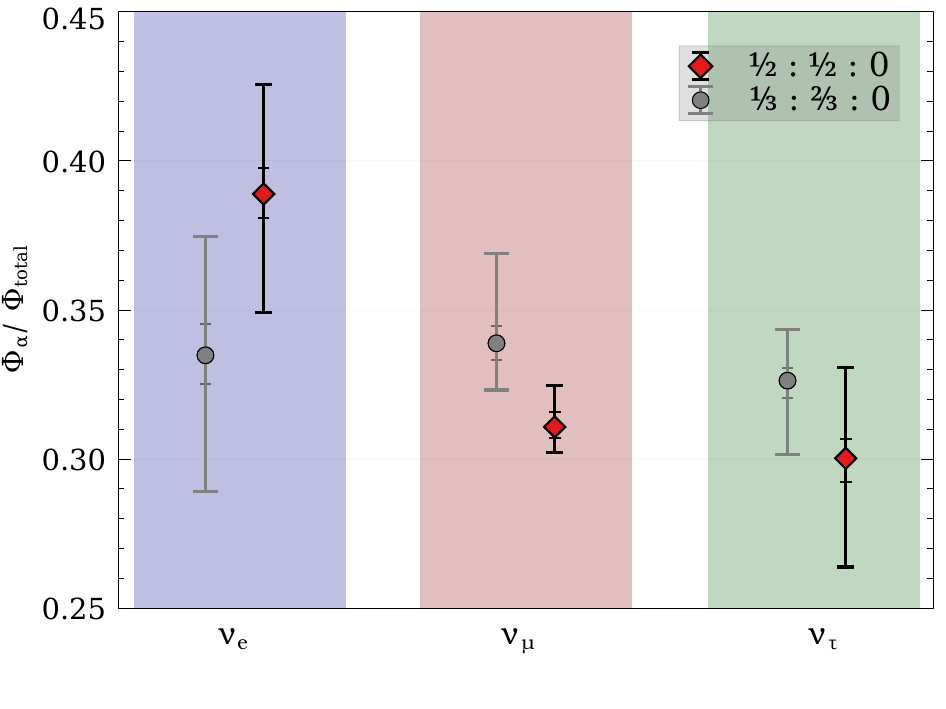}
	\caption{\label{fig:flavfractions} Flavour compositions as
		fraction of total flux at Earth for best fit mixing parameters and
		3$\sigma$ variation thereof.  The smaller uncertainties  are obtained
		using  projected 2040 uncertainties on the underlying mixing parameters,
		with results especially from JUNO constraining $\sin^2(\tht{12})
		$~\cite{JUNO:2015zny,JUNO:2022mxj} while results from
		Hyper-Kamiokande~\cite{Hyper-Kamiokande:2018ofw}
		and DUNE~\cite{DUNE:2020lwj} do the same to $\tht{23}$ and to
		a lesser extent to \dcp.}
\end{figure}

For a flux with its origins in heavy quark decays that starts with $\onehalf :
	\onehalf : 0$ at the source ((\onehalf :
	\onehalf : 0)-origin), we find that neutrino flavour mixing during
propagation results in a composition at Earth, including mixing parameter uncertainties, of ${(0.39 \pm 0.04 : 0.31 \pm 0.01 :
			0.30^{+0.03}_{-0.04})}_\oplus$. % including mixing parameter uncertainties.
For comparison, for a flux at the source with a normalised flavour composition
of $\onethird : \twothird : 0$ (e.g.~from pion decay), it is well-known that
the best-fit flavour fractions at Earth are nearly $\onethird: \onethird :
	\onethird$, or more precisely ${(0.33 \pm 0.04 : 0.34^{+0.03}_{-0.02} :
			0.33^{+0.02}_{-0.03})}_\oplus$.

In Fig.~\ref{fig:flavfractions}, we demonstrate these results for
the three flavour fractions, including their variation due to current
mixing matrix parameter uncertainties.
With current uncertainties,
resolving the two different flavour compositions considered here is a challenging
prospect for any experiment.
On the other hand, with future measurements of these parameters at dedicated neutrino
oscillation experiments constraining them more stringently, distinguishing between
individual flavour fractions may potentially become feasible by the year 2040.

\begin{figure}
	\centering
	\includegraphics[width=0.75\textwidth]{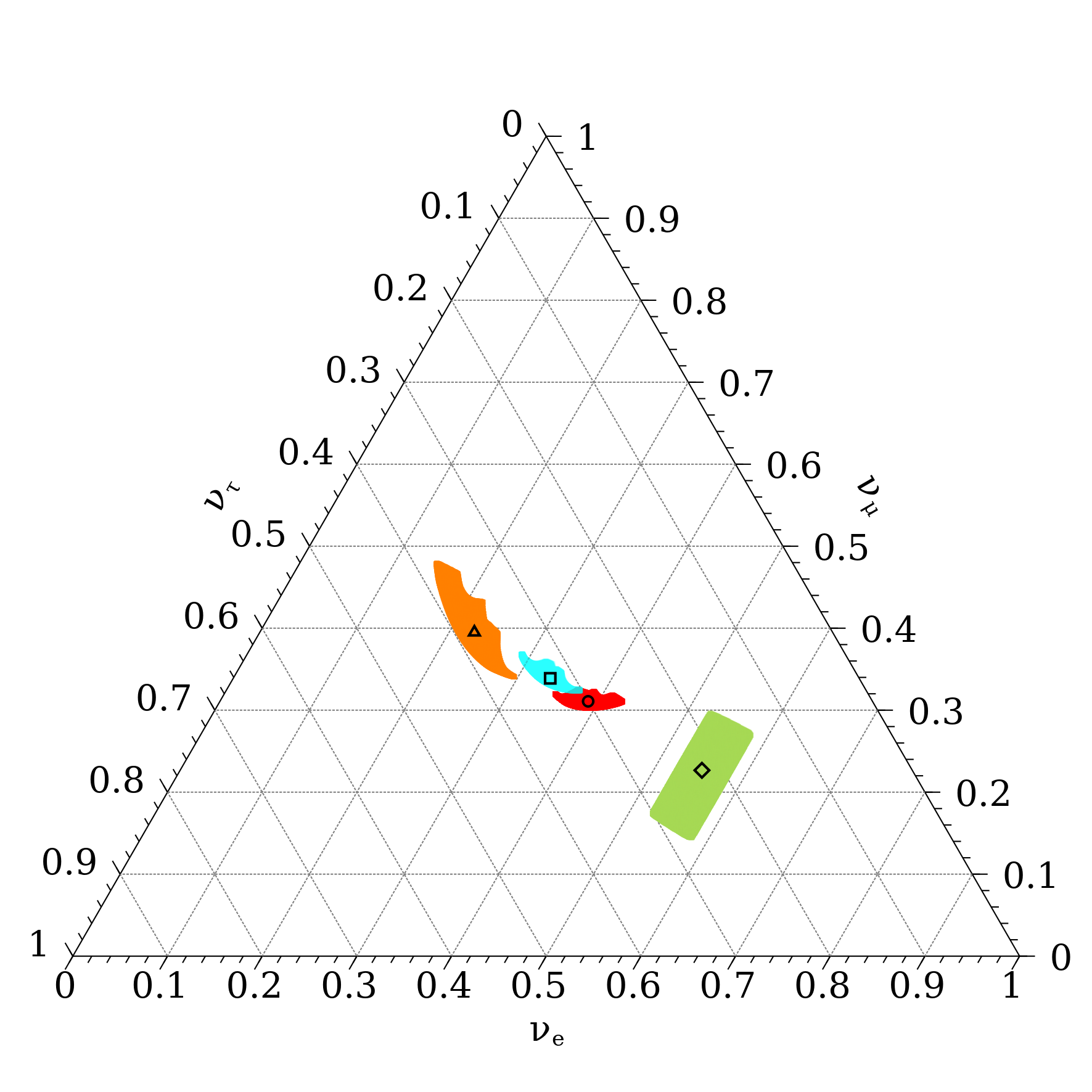}
	\caption{\label{fig:flavtriangle} Flavour fractions at Earth for best-fit and current
		3$\sigma$ variation of mixing matrix parameters computed with various
		initial flux fractions.  The red patch with $\circ$ showing the best-fit
		point is for a starting flavour composition of $\onehalf :\onehalf :0$
		(charmed meson decays, this work, and {\bf  see also \cite{Bustamante:2015waa}}), while the cyan patch with $\square$ as
		best-fit point is computed assuming a starting flux of $\onethird :\twothird
			:0$~\cite{Song:2020nfh}. The green patch ($\diamond$) and orange
		($\vartriangle$) patches assume neutrino fluxes originating from from neutron
		decays (initial composition $1:0:0$, for example for cosmogenic fluxes) and
		muon-damped sources ($0:1:0$ starting flavour)
		respectively~\cite{Song:2020nfh}. }
\end{figure}

An alternative way to demonstrate the flavour composition variation at Earth
given an initial composition at source is in terms of the flavour fraction
triangle as shown Fig.~\ref{fig:flavtriangle}.
Thus, given a normalised flavour composition of $\onehalf : \onehalf : 0$
at source, the
red patch therein shows potential flavour fractions at Earth for all possible
variations of the mixing matrix parameters within their current 3$\sigma$
uncertainties.
This contrasts with the cyan patch which is similarly computed for a starting
composition of $\onethird : \twothird : 0$.

Translated into flavour fraction ratios at Earth, we find that the $\Phi_{\nu_e}
	/ \Phi_{\numu}$ ratio is $\approx 1.25$ for a starting composition of
$\onehalf :\onehalf :0$ as opposed to $0.99$ for a $\onethird :\twothird :0$ initial flux.
Concomitantly, the $\nutau/\nue$ flavour ratio changes from 0.77 to 0.98,
while, on the contrary, the $\numu/\nutau$ ratio remains largely unchanged.

\begin{figure}[htb]
	\centering
	\includegraphics[width=0.49\textwidth]{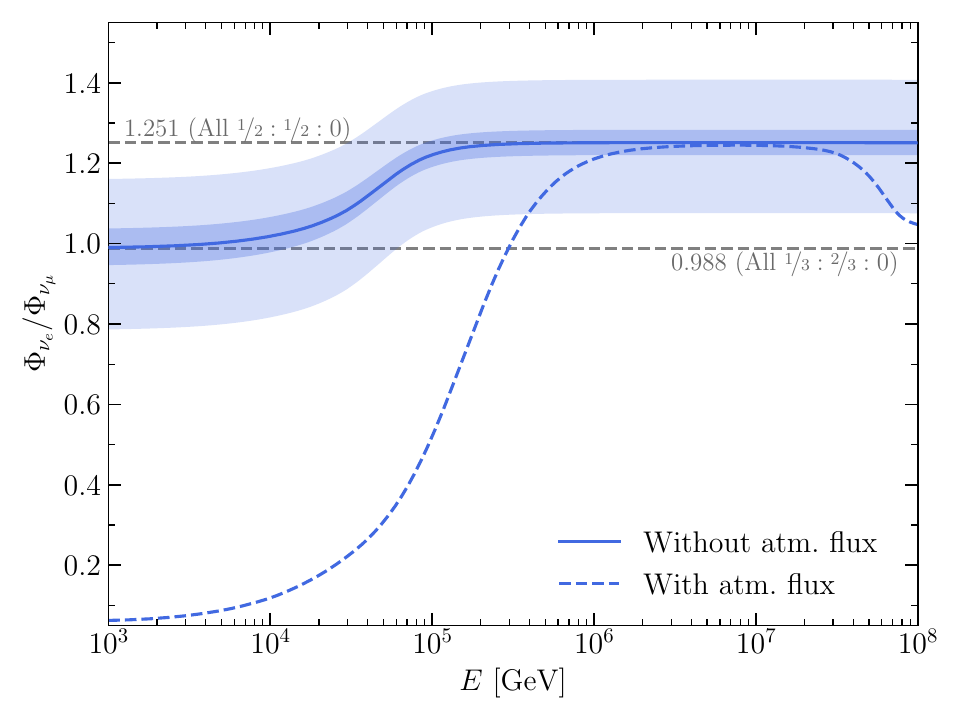}
	\includegraphics[width=0.49\textwidth]{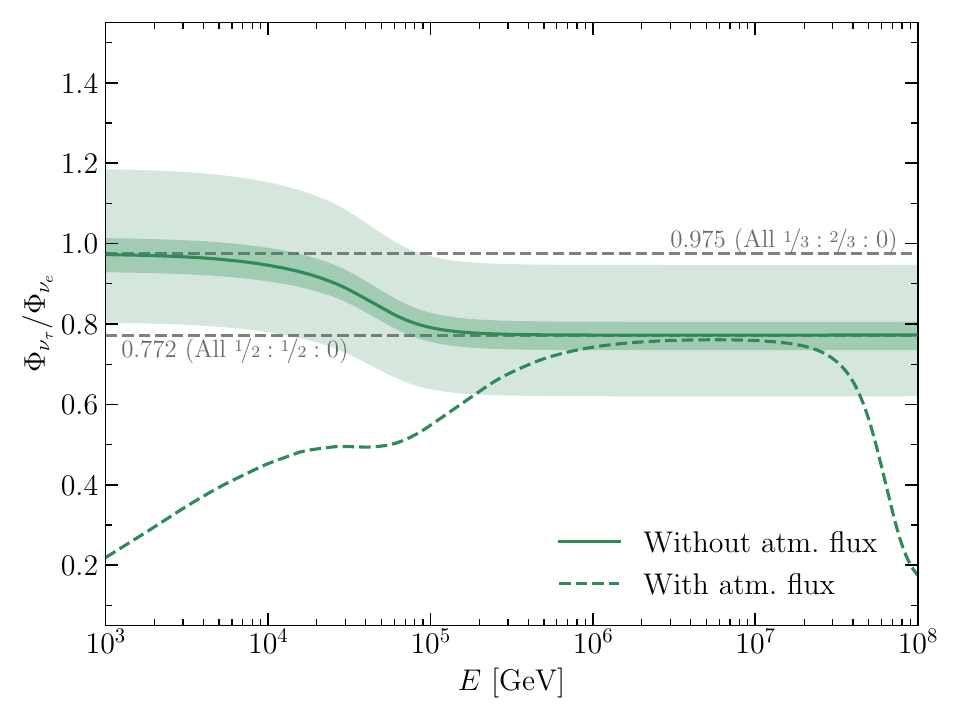}
	\\
	\includegraphics[width=0.49\textwidth]{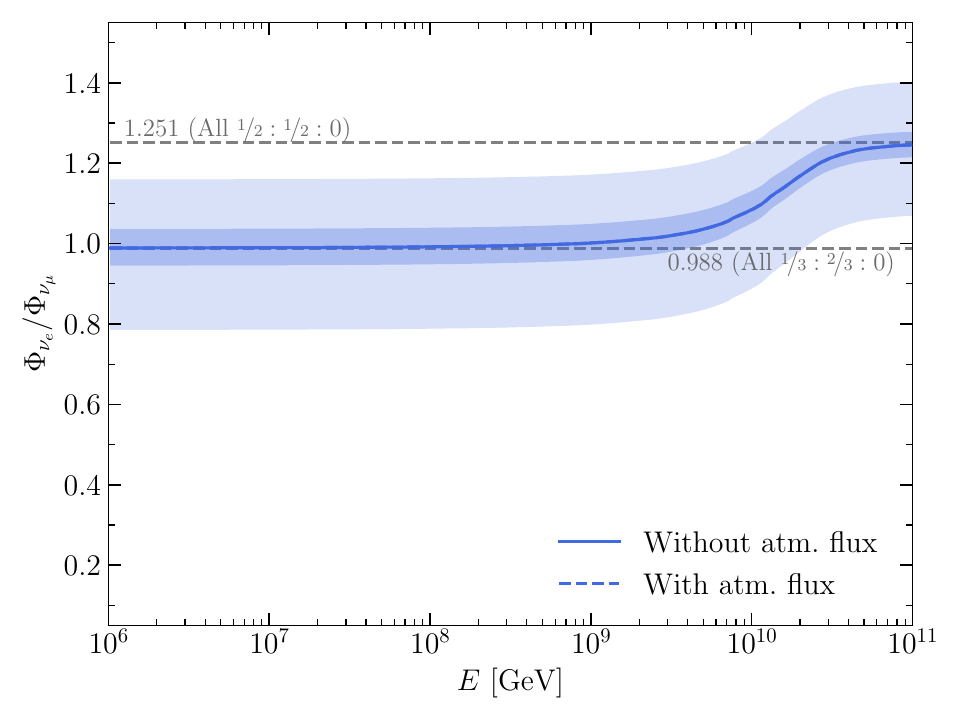}
	\includegraphics[width=0.49\textwidth]{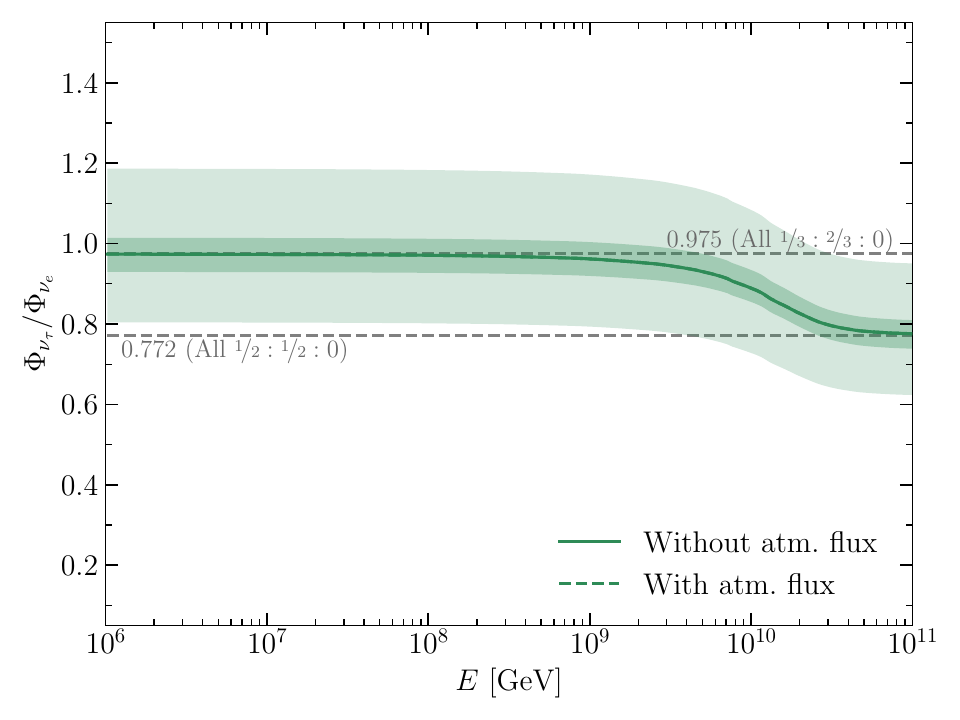}
	\caption{\label{fig:flvar} Energy dependent flavour ratios at Earth
		for our sample SJS (MdSn) source shown in the top (bottom) panel.
		Uncertainty bands are computed using current
		3$\sigma$ ranges of the mixing matrix parameters deduced while including SK
		atmospheric data~\cite{Esteban:2020cvm}.
		Narrower uncertainty bands are derived assuming reduced uncertainties
		on mixing parameters projected by the year 2040.
		For SJS sources, the dip in the ratios seen at high energies $\gtrsim 30$
		PeV is due to the source flux dropping exponentially beyond a few PeV and
		quickly falling below atmospheric flux levels (see Fig.~\ref{fig:fluxes}).
	}
\end{figure}

Where two fluxes with different starting flavour compositions dominate at
different energies, such transitions in the flavour ratios will be seen to
occur at a specific energy.
Specifically, for the case of SJS sources described previously, neutrinos
from kaon-decays comprise the dominant component of flux up to energies of
$\sim 100$ TeV and those from decays of charmed mesons likewise at higher
energies; one therefore expects to find the $\nue/\numu$ flavour ratio to
show a transition from 0.99 to 1.25 close to these energies.
We may therefore define an energy dependent ratio between any two flavour
components (say, $\alpha$ and $\beta$) of the total neutrino flux across the entire
energy range as
\begin{equation}
	\mathcal{F}_{\alpha\beta}(E_\nu) = \frac{\sum_{i} \Phi^{i}_\alpha(E_\nu)}
	{\sum_{i}\Phi^{i}_\beta(E_\nu)}\,,
	\label{eq:flrenergy}
\end{equation}
where the summation over $i$ iterates over fluxes \emph{at Earth} from
every relevant source.
For example, in the SJS case, this summation would include the atmospheric
neutrino flux of the specified flavour and fluxes from SJS due to kaon and
charmed meson decays.
As shown in Fig.~\ref{fig:fluxes}, the dominant component of the flux changes
over the energy range, in this case, from atmospheric neutrinos up to tens of
TeV, to those from kaon decays at SJS up to 100 TeV, and finally to the flux
from charmed meson decays.
This induces a non-trivial energy-dependence on the flavour ratios as shown in
Fig.~\ref{fig:flvar}.
However, the dominance of the atmospheric neutrino flux over both components of
the SJS flux up to energies of several tens of TeV, alters the picture
significantly.
Since this flux is dominantly comprised of muon neutrinos at low energies
(see Fig.~\ref{fig:fluxes}), we find that the $\nue/\numu$ flavour ratio drops
steeply at energies below $\sim 50$--60 TeV until it falls to about $0.05$ at
1 TeV.
Conversely, the rise of this ratio with energy, first to $\approx 1$ at energies
of around 250 TeV, where the atmospheric neutrino flux is all but gone, and thereafter to
$1.25$ at higher energies is markedly steeper than it would be were the
transition to solely involve the two SJS fluxes.

Similar conclusions may also be drawn for the $\nutau/\nue$ flavour ratio,
which, in the absence of atmospheric neutrino fluxes, should drop from a
value of 0.975 at the lower energies, up to 100 TeV, to 0.772 thereafter.
However, in this case, the dominance of atmospheric neutrino fluxes ---
with very little \nutau\ content --- at lower energies completely changes
the nature of this transition: the
$\nutau/\nue$ flavour ratio rises, instead of falling, from lower to higher energies
until, at around a PeV, it reaches a value of 0.772 at $\sim 1$ PeV marking the
beginning of the dominance of the neutrino flux from charmed meson decays.
We show these results, including the uncertainties associated with the flavour
ratios, in Fig.~\ref{fig:flvar} for SJS sources (top panel).

For MdSn sources, similar transitions appear at much higher energies $\sim 10$
EeV, making the consideration of atmospheric neutrino fluxes moot.
The flavour ratio as described in equation~\ref{eq:flrenergy}, therefore, only
involves astrophysical fluxes.
We show these results in the bottom panel of Fig.~\ref{fig:flvar}.

\begin{figure}[htb]
	\centering
	\includegraphics[width=0.95\textwidth]{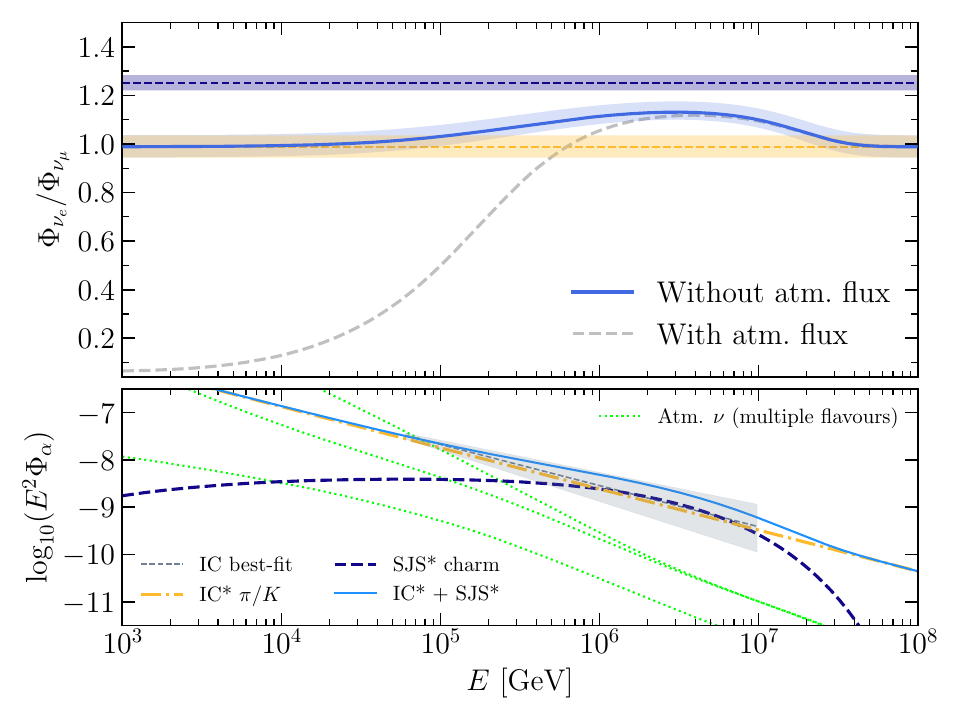}
	\caption{\label{fig:flvarscaled} Top: The $\Phi_{\nu_e} / \Phi_{\nu_\mu}$
		flavour ratio (solid blue curve) as a function of energy for the combined
		power-law and SJS-charm flux shown in the lower panel, without the
		angle-averaged atmospheric neutrino flux contribution. The dashed grey line
		includes the atmospheric neutrino flux. For comparison, flavour ratios
		corresponding to a fully $\left( \onethird: \twothird:0 \right)$-origin
		flux (light orange) and fully $\left( \onehalf: \onehalf : 0\right)$-origin
		flux (purple) are shown along with their $3\sigma$ maximum and minimum
		ranges allowed by the projected 2040 uncertainties of mixing parameters.
		\textbf{Bottom:} The different fluxes used in this analysis. Fluxes
	  labelled with an `$*$' have been rescaled as described in the text.  For
	  comparison, the unscaled central value of the IceCube HESE best-fit single
	  power law curve is shown with grey dashes, along with its uncertainty band in
	  grey.  The total flux comprising the rescaled power-law and the SJS-charm flux
	  lies within the uncertainty band corresponding to the IceCube best-fit
	  power-law flux between the energies of 60 TeV--10 PeV. The three flavours of
	  atmospheric neutrinos are shown as light green dotted curves.}
\end{figure}

With our choices of parameters for the population of SJS, we find that the
charm-origin neutrino flux completely dominates over other flux components
between 1--10 PeV energies, however, more realistic scenarios may lead to fluxes from
SJS that constitute only a fraction of the total flux admixture, with the remainder
coming from $\pi/K$ decays.
Current uncertainties in the neutrino mixing parameters preclude the
possibility of disentangling a fully $\left( \onehalf:\onehalf:0
	\right)$-origin flux from a fully $\left( \onethird:\twothird:0
	\right)$-origin flux.
By contrast, the significant reduction in uncertainties projected by 2040
offer hope of detecting similar flavour ratio transitions.
It also raises the question as to what fraction of the total flux reaching
Earth needs to be of a charm origin --- assuming the remaining flux to
originate with a $\onethird:\twothird:0$ composition --- to make the
transition discernible accounting for the 2040 mixing parameter
uncertainties.\footnote{In practice, any detectibility would depend on the
event statistics and therefore hinge critically on the energy where the
transition starts.}
To understand this, we analyse a hypothetical scenario in which the flux from
SJS is reduced by a constant factor and combined with the IceCube single
power-law flux. With the power-law flux normalized to 85\% of the best-fit
value, labelled as IC*, still within the IceCube uncertainty band of the flux
normalization~\cite{IceCube:2020wum}, the SJS neutrino spectrum is normalized
(labelled as SJS*) such that at neutrino energy of 1 PeV,  the SJS* and IC* are
equal. The combined flux 
lies within the uncertainty band corresponding to the IceCube best-fit power-law
flux between energies of 60 TeV--10 PeV, as shown in the lower panel of
Fig.~\ref{fig:flvarscaled}.
Assuming the IceCube power-law spectrum comes fully from $\pi/K$-origins,
practically, this implies that the charm-origin flux component in the
total flux would be 50\%.
We use this total flux to investigate how the flavour ratios change as a
function of energy.
In particular we show the $\Phi_{\nu_e} / \Phi_{\nu_\mu}$ flavour ratio in the
upper panel of Fig.~\ref{fig:flvarscaled}.
We note that that while the maximum flavour ratio in this scenario does
not rise to a value of 1.251 representative of a dominant charm-origin flux,
the central value grows gradually to a maximum of about 1.131 (1.119 when
including atmospheric neutrino fluxes in the total flux) --- higher than the
maximum value of the ratio for a fully $\left( \onethird: \twothird: 0
	\right)$-origin flux allowed by the $3\sigma$ projected 2040 uncertainty
ranges of mixing parameters.
The rise becomes tangible at energies between 300--400 TeV.

\section{\label{sec:ic}Inferring flavour ratio changes from event-rate ratios
  of different morphologies}%
%!TEXroot = ./main.tex

Events originating from neutrino-nucleon ($\nu_\alpha N$) interactions, where the
subscript $\alpha$ refers to any of the three neutrino flavours, occurring
within IceCube's instrumented volume may be classed into different groups based
on their morphologies that depend on the initiating neutrino flavour and the
mediating boson:
\begin{enumerate}
	\item $Z^0$-boson mediated neutral current event initiated by any
	      flavour produce cascades, with a final state neutrino carrying away
	      energy from the interaction without detection by the detector. For
	      these events, the energy deposited in the detector, \edep, is
	      substantially different from the energy of the initial neutrino
	      $E_\nu$.
	\item Charged current events initiated by $\nu_e N \to e^{-} X$
	      interactions are detected as nearly-spherical cascades in the detector.
	      As opposed to neutral current events, the entire energy from the
	      incoming $\nu_e$ is deposited in the cascade with the outgoing $e^-$
	      almost immediately losing its energy as well; thus, in this case we can
	      assume $\edep = \enu$.
	\item On the contrary, a charged current event from $\nu_\mu N \to \nu^- X$
	      produces a cascade, with substantial energy taken away by the
	      final state muon. As the muon scatters away from the interaction, it
	      loses energy predominantly via bremsstrahlung, with the resulting
	      track-like event topology a signature of a \numu\ initiated event.
	\item Charged current events initiated by $\nu_\tau N$ interactions also
	      show up as cascades; however, the outgoing $\tau^{-}$ shows up as
	      a faint track that traverses a short distance before it decays
				producing a secondary cascade or ``bang''.
	      If both cascades are detected within the detector's instrumented volume
	      but separated by enough distance to allow the cascades to be
	      individually resolved, this ``double-bang'' becomes a smoking gun
	      signature of a charged current \nutau\ event.
\end{enumerate}

With long-lived tracks being exclusive to $\numu$ initiated events,
estimates of the cascade to track event rate ratios~\cite{Song:2020nfh}
across the entire HESE-sensitive span of energies will throw light on the
relative \numu\ content in the neutrino flux arriving at Earth.
In turn, any change in this quantity would directly point to
changes in the flavour content of the dominant flux between 10 TeV and 10 PeV.
Having already computed the final flavour composition at Earth for both
pion/kaon-dominated initial fluxes (flavour fraction $\onethird :\twothird :0$
at source) and heavy meson dominated initial flux ($\onehalf :\onehalf :0$
initial flavour composition), we can translate these results into event ratios
between different morphologies by making use of deeply inelastic $\nu N$
scattering cross-sections in the literature~\cite{Cooper-Sarkar:2011jtt}.

\begin{figure}[htb]
	\centering
	\includegraphics[width=0.75\textwidth]{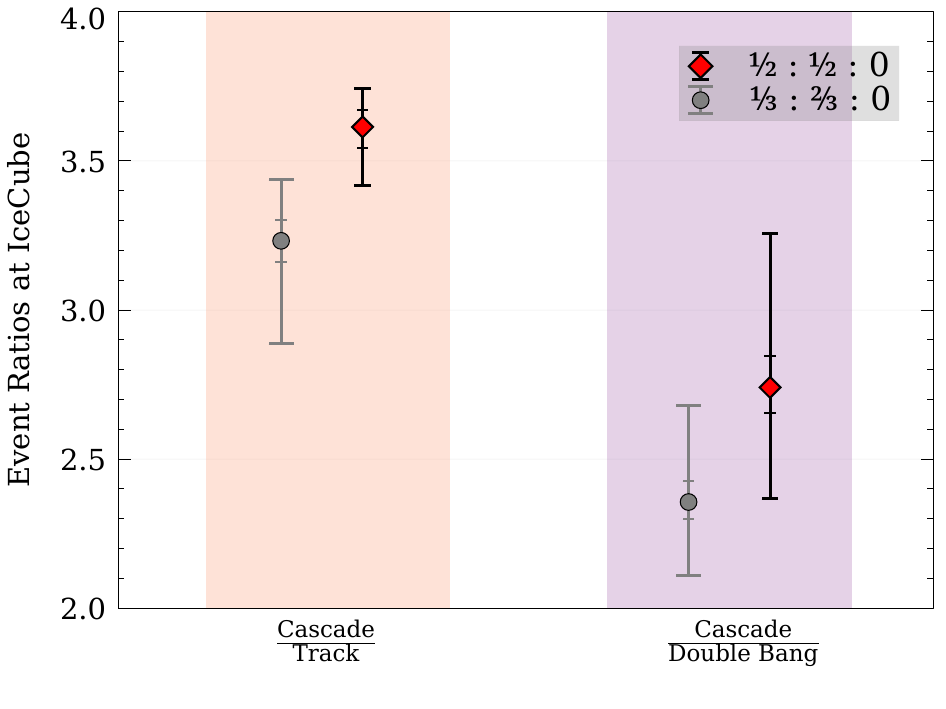}
	\caption{\label{fig:eventratios}Cascade to track and cascade to double
		bang event ratios at Earth detectors for fluxes starting at $\onehalf
			:\onehalf:0$ and $\onethird :\twothird :0$ at the source.
		The evaluation is made at energies of 10 PeV, at and over which \nutau\
		signature events are expected to be distinguishable from general cascades.
		Error bars represent variation in the ratios due to $3\sigma$ uncertainties
		of mixing parameters, both current (bigger error bars) and projected for
		2040 (smaller).}
\end{figure}

In Fig.~\ref{fig:eventratios}, we show the best-fit cascade to track
ratio for a neutrino flux originating with an initial composition of
$\onethird:\twothird :0$ and compare that with an initial composition of
$1:1:0$.
We find that for a neutrino flux originating from charmed meson decays in the
composition $\onehalf :\onehalf :0$, the expected cascade-to-track event ratio is
$3.61^{+0.13}_{-0.19}$, as opposed to $3.23^{+0.20}_{-0.34}$ for a $(\onethird
	:\twothird :0)$-origin flux (e.g.,\ from pion decays).
The uncertainties represent the maximum and minimum values of the ratios,
obtained in each case by varying the mixing matrix parameters over their
current 3$\sigma$ range.
With projected uncertainties in these parameters expected to
be significantly reduced in the future, we find that cascade-to-track ratios
will allow resolving the two different initial flux compositions at more than
$3\sigma$ significance by the year 2040.
Similarly, we find that at high energies, assuming the efficiency of resolving
tau double bangs from other cascades becomes 100\% efficient, the cascade to
double bang ratio becomes $2.74^{+0.49}_{-0.37}$ for a flux with a $\onehalf
	:\onehalf :0$ starting flavour composition, in contrast to
$2.36^{+0.32}_{-0.25}$ for one starting with $\onethird :\twothird :0$.

The transition from pion dominated to charm happens at a specific
energy (depending on conditions in source).
As we have seen in Fig.~\ref{fig:flvar}, this induces a non-trivial energy
dependence in flavour ratios which directly carries over to cascade-to-track
event ratios; additionally, the latter are also influenced by a mild
$E$-dependence of the ratio of CC to NC cross-sections.

For \nutau\ initiated events, the higher the energy of the initial \nutau\,
the longer the out-going $\tau$ track.
Towards the lower-energy end of the HESE data, relatively short-lived tau
tracks imply the final $\tau$ decay cascade cannot be resolved from the more
catastrophic initial cascade which either completely or partially overlaps the
former.
For incoming \nutau\ with super-PeV energies, on the other hand, the final tau
becomes sufficiently long-lived that the two bangs can be distinguished from
each other and $\tau$ double bangs become an exclusive class of events.
Thus, while IceCube can distinguish between cascades including double-bangs and
tracks across the entire energy range, it is only weakly sensitive to
tau-specific signatures at lower energies ($\lesssim 1$ PeV).
As a consequence, double bang/cascade or double bang/track
become useful only at high energies.
However, this means that the transition in the flavour ratio needs to
happen at energies beyond a PeV.
This is not the case for SJS sources we consider but may work for other
sources.
As an example, for magnetars, the kaon-charm transition occurs at
$\sim 10^{10}$ GeV~\cite{Carpio:2020wzg}, however poor statistics make it
challenging to draw any conclusions.

\begin{figure}[htb]
	\centering
	\includegraphics[width=0.85\textwidth]{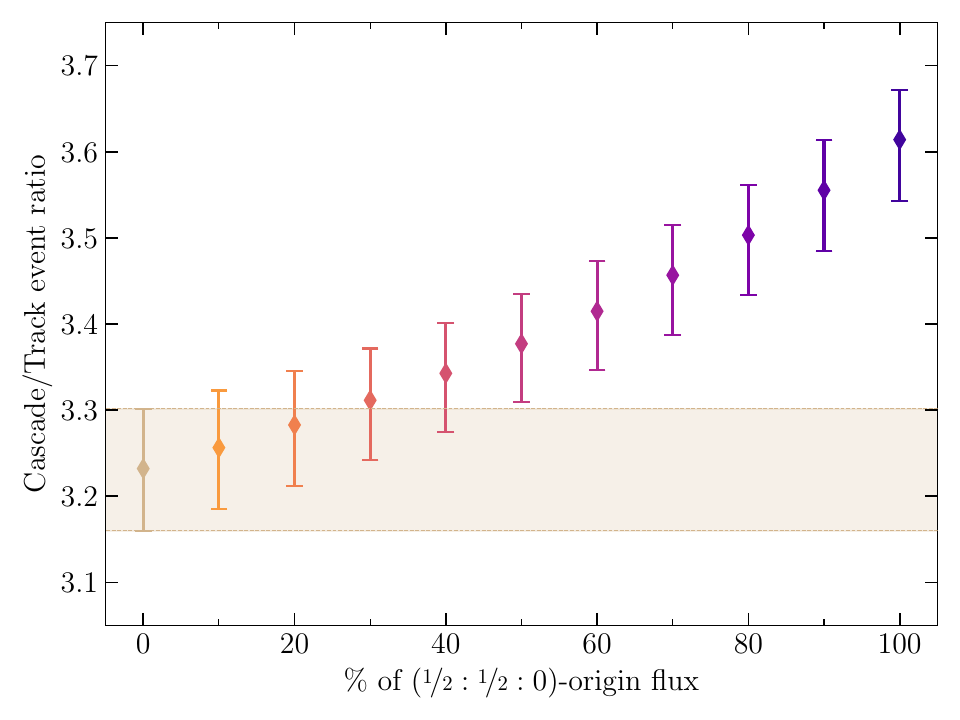}
	\caption{\label{fig:csc2trk-frac}Cascade to track event ratios as a
		function of the percentage flux coming from charm-decay with the rest
		assumed to originate with a $\onethird:\twothird:0$ composition. For each
		point, the error bars represent the uncertainty of the ratio for a full
		scan of the neutrino mixing parameters over their projected 2040 3$\sigma$
		uncertainties. We show this uncertainty band corresponding to the fully
		$\onethird:\twothird:0$ flux extended across the entire $x$-axis (tan band)
		to illustrate what percentage of $\onehalf:\onehalf:0$ flux in the
		admixture causes the ratio to become distinguishable from this band.}
\end{figure}

As discussed above, an admixture of neutrinos from $\pi/K$ and an appropriately
normalized SJS spectrum of neutrinos that is dominated by charm is consistent
with IceCube measurements of the diffuse flux. To display the sensitivity of
the cascade to track ratio to an admixture,  in Fig.~\ref{fig:csc2trk-frac} we
show the cascade to track event ratio as a function of the percentage of
charm-origin flux in the admixture. As usual, the remainder is assumed to be
composed entirely of neutrinos from $(\onethird: \twothird: 0)$-origin. For
each admixture, the upper and lower limits represent variation of the mixing
parameters over their 3$\sigma$ 2040 projected uncertainties.  We find that at
least a $\sim 50\%$ proportion of the neutrino flux needs to come from charm
decays in the source to be discernible from neutrino sources dominated by
$\pi/K$ decays.

\subsection{\label{ssec:csc2trkHESE}Analysis of cascade to track ratio using IceCube HESE 7.5yr data}

In this section, we investigate if current data, notwithstanding
the low statistics, show any trend of transition in flavour ratios within
10 TeV to 10 PeV.
To this end, we make use of the IceCube HESE data-set collected over 7.5
years.
To the extent that IceCube can clearly distinguish between cascades and
$\mu$-tracks, over the entire HESE energy range but can only distinguish double
bangs from \nue\ cascades at the higher energy end of the spectrum, for our
analysis, we shall categorise double bangs and \nue\ cascades together distinct
from \numu\ CC tracks.

The HESE sampling comprises 102 events with \edep\ measured between 10 TeV--2.1
PeV~\cite{IceCube:2020wum}.
The lower-end of the spectrum ($\edep \sim [10, 60]$ TeV) is dominated by
atmospheric neutrino fluxes, despite the analysis involving significant
background rejection.
Instead, we focus on the 60 events with deposited energies above 60 TeV,
allowing us to work with a pre-dominantly extra-galactic flux throughout.

While the data identifies each event with its \edep\ and morphology, we need to
``re-construct'' the incoming neutrino energy for each.
We do this using a simple Monte-Carlo analysis, whereby, out of the entire set
of cascade events, we randomly categorise events as CC or NC based on the
ratio of neutral-current to charged-current $\nu N$ cross-sections.
For the former, we assume $\edep \approx \enu$;
for the latter we scale the \edep\ to get \enu: $E_\nu = \edep / \langle
	y(E_\nu) \rangle$, where $y$ is the inelasticity of the interaction.
At these energies, $\mean{y(E)}$ is slowly-varying, so we simplify the
analysis by assuming $\mean{y(E_\nu)} \approx \mean{y(\edep)}$.
At the end of this exercise, we are left with a data-set of 60 events, with
each identified with its initial neutrino energy \enu\ and cascade or track
morphology.

\begin{figure}
	\centering
	\includegraphics[width=0.49\textwidth]{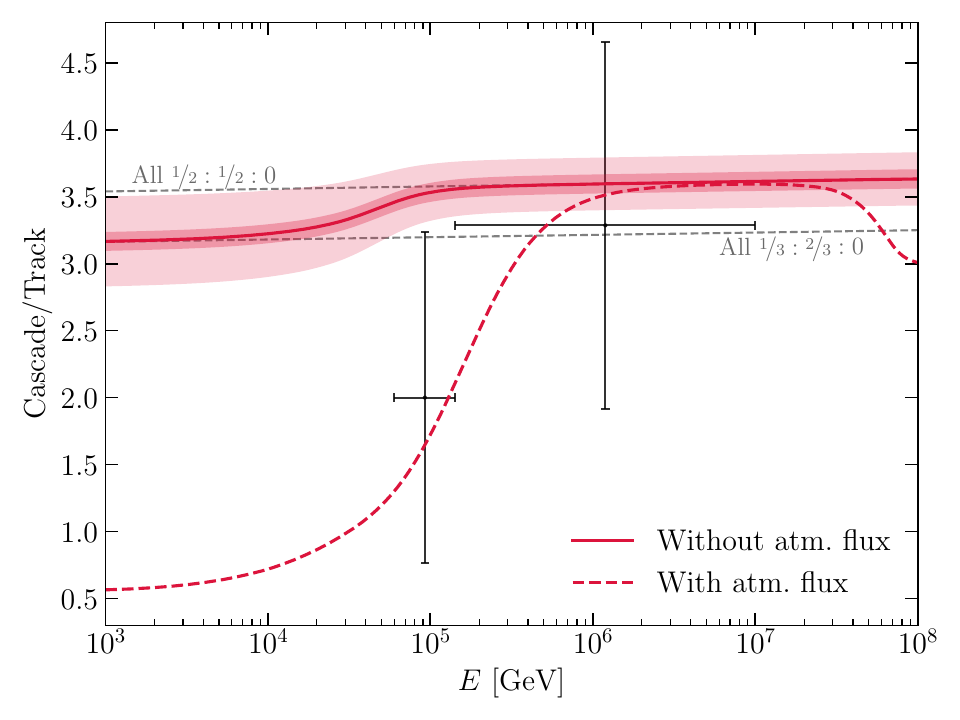}
	\includegraphics[width=0.49\textwidth]{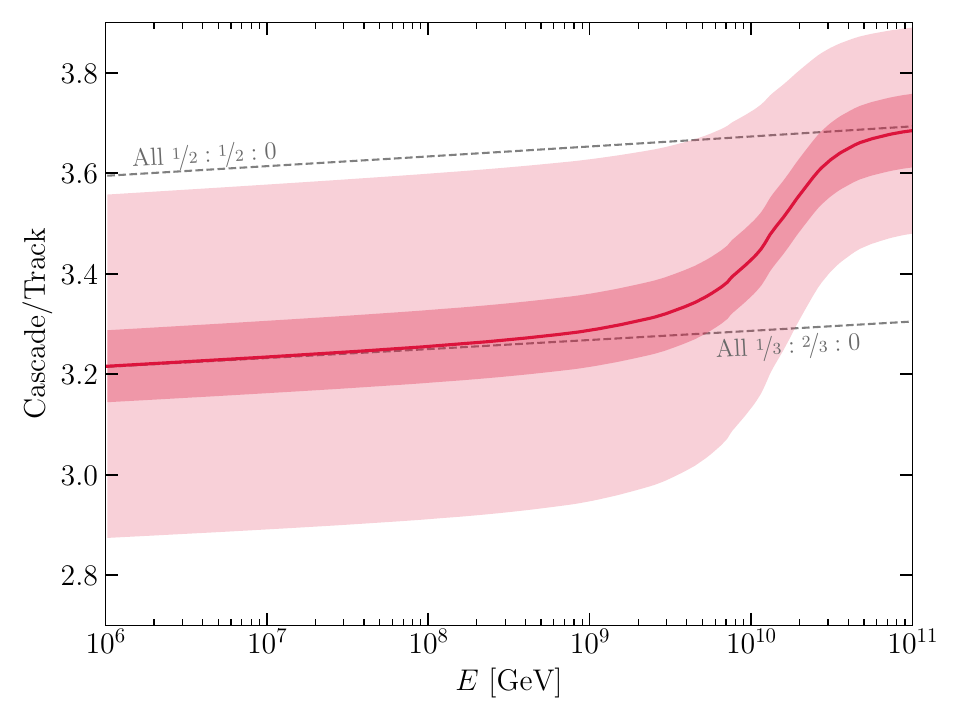}
	\caption{\label{fig:SJSC2T}\textbf{Left:} Ratio of cascade to track events as
		a function of neutrino energy $E$ assuming the extra-galactic flux to
		originate at slow-jet sources. The steep drop off of the ratio at lower
		energies is due to the dominance of atmospheric neutrino fluxes.
		Also shown are the ratios inferred from IceCube HESE 7.5-yr data at two
		separate bins as described in the text.
		The grey error bars represent $1\sigma$ Poisson uncertainties
		on the mean value of the ratio for each bin.
		\textbf{Right:} Same as left, but for MdSn sources. Note that, since
		atmospheric fluxes play a negligible role at PeV energies and higher, we
		have omitted them from this plot, in contrast to the SJS plot on the left.
		In both figures, uncertainty bands are computed using 3$\sigma$ uncertainty
  	ranges of the current (light bands) and 2040 (darker bands) mixing matrix
	  parameters.}
\end{figure}

Since our goal is to determine flavour ratio transitions between low and high
energies, we bin the events by their \enu.
To assign equivalent statistical weights to low and high energy bins, we split
the set of events into two bins with an equal number of events (i.e.\ 30) in
each.
Note that, since event statistics worsens with increasing energies, binning in
this manner leads to the bin widths being very unequal.
We can now compute the ratio of the number
of cascades and double bangs taken together to the number of muon-track events
in each bin.
To offset the randomness of classification of cascades into NC and CC events,
we run the MC analysis over a million times and extract the mean ratios for the
two bins.
The results are shown in the left panel of Fig.~\ref{fig:SJSC2T}, as compared
with the flavour ratio transition expected for the case of SJS sources.
We find that the IceCube ratio as evaluated here changes from $2.00 \pm 1.24$ for the lower energy bin to
$3.39 \pm 1.39$.
The error bars on the ratios represent $1\sigma$ Poisson uncertainties
around the mean values.
We specifically emphasise that these uncertainties do not directly use or
correlate to the uncertainties on neutrino mixing matrix parameters.

While the computed cascade-to-track ratios show hints of growth at
higher energies, the associated uncertainties --- a consequence of the
inevitable low event statistics --- are simply too large to allow a definitive
conclusion whether there is a growth that %as to whether the growth specifically 
points towards the dominance
of a $(\onehalf :\onehalf :0)$-origin flux at the higher energies. Certainly,
with current statistics,  a  $50:50$ mix of $\pi/K$ and charm sources at $E_\nu
=1$ PeV illustrated in Fig.~\ref{fig:flvarscaled} cannot be tested.
More fine-grained binning would improve the energy resolution of the flavour
transition; however this would be at the cost of even poorer per-bin statistics,
in turn leading to even higher error bars on the final ratios.
Nonetheless, we have checked that the trend of growth in the ratio persists
even with finer binning, by using 3 (20 events / bin) and 4 bins (15 events /
bin), although as mentioned before, it becomes progressively weaker with poorer
per-bin event statistics.
Low statistics of current PeV-scale and higher energy data, with the HESE data containing only 3
events with deposited energies over a PeV, prohibit a similar analysis
involving MdSn sources.

We note that the analysis presented in this section does not take into
account experimental efficiencies in distinguishing between the different
event morphologies, nor any particular experimental biases.
It serves as a proof of principle for how similar analyses, potentially with refinements
incorporating such considerations, may be carried out to determine
energy-dependent flavour ratio variations in the future when more statistics will have
accrued at IceCube or other neutrino telescopes.

\section{\label{sec:conclu}Summary and conclusions}%
%!TEXroot = ./main.tex
The many successes of IceCube since starting taking data in 2010 as an
86-string, 1 km$^3$ neutrino observatory have opened up an exciting window into the
study of highly energetic astrophysical sources where high energy
neutrinos originate and the propagation of these elusive particles.
Understanding the flavour composition of the incoming neutrino flux is
a key component towards understanding the nature of its source, complementary
to and as important as understanding the overall spectral shape and magnitude
thereof.

While the question of flavour fractions arriving at Earth has been discussed
in the literature, such discussions have typically assumed starting flavour ratios
of $\frac{1}{3}:\frac{2}{3}:0$ from the decay of light mesons, or 1:0:0 from neutron decay, or
0:1:0 from muon damped sources.
In this work, we have focused on charmed meson decay initiated neutrino fluxes
that lead to a $\frac{1}{2}:\frac{1}{2}:0$ distribution of flavours at source.
Using current best-fits for neutrino mixing parameters, we have shown
that standard flavour mixing reduces this to $0.39:0.31:0.30$ by the time the flux
reaches Earth, as opposed to approximately
$\frac{1}{3}:\frac{1}{3}:\frac{1}{3}$ at Earth for a
$\frac{1}{3}:\frac{2}{3}:0$ initial flavour composition.
Using this result and uncertainties associated with neutrino flavour mixing parameters,
we have updated the corresponding patch in the flavour triangle, see Fig.~\ref{fig:flavtriangle}.

%Additionally, 
We have discussed the implications for IceCube should the
dominant neutrino flux differ in its flavour composition at lower (TeV) energies
vis-à-vis at higher (PeV) energies.
Within the realm of standard model physics, this could happen if neutrino
production at source is predominantly from light meson decays at one edge
of the spectrum, but from charmed and heavier mesons at the other.
This would be the case for neutrino fluxes originating from slow-jet sources
and magnetar-driven supernovae, for instance.
For the former, we have shown that for very representative assumptions of
source parameters, the transition in flavour composition
may happen within the TeV-PeV energy range to which current IceCube data is sensitive.
For the latter, a similar transition is expected at significantly
higher energies (between $10^9$--$10^{11}$ GeV), with low event statistics
at these energies making the transition significantly more challenging to
detect. Radio detection of neutrinos~\cite{Ackermann:2022rqc} in this
ultra-high energy range might be able contribute to flavour
studies~\cite{Glaser:2021hfi}.

Knowing that the transition may be quantified in terms of flavour fraction ratios, and
concretely seen in neutrino telescope experiments in terms of ratios of
different event morphologies, we have used a sample of current IceCube data --- HESE 7.5-yr
--- to look for any hints pointing to such transitions.
The IceCube best-fit single power law with uncertainties permits a $50:50$
admixture of a power law from $\pi/K$ decays and a SJS charm contribution to
the diffuse astrophysical neutrino flux.
We find that the relatively low statistics of current events is insufficient to allow 
a conclusion about the flavour composition at sources, but we expect that the
method used here, scaled up for data from \Gen, will provide more definitive
conclusions.

\acknowledgments%
We thank M.~Bustamante, C.~P\'erez de los Heros, C.~Glaser and A.~Hallgren for
discussions.  This work is supported in part by the U.S.~Department of Energy
Grant DE-SC-0010113 (MHR) and in part by the U.S.~Department of Energy Grant
DE-FG02-13ER41976/DE-SC-0009913 (IS).

\bibliographystyle{JHEP}

\bibliography{sample}

\end{document}